\documentclass{article}
\usepackage[utf8]{inputenc}
\usepackage{verbatim}
\usepackage{amssymb}
\usepackage{amsmath}
\usepackage{mathrsfs}
\usepackage{mathtools}
\usepackage{color}
\usepackage{amsthm}
\usepackage{fullpage}
\usepackage{enumerate}

\usepackage{titlesec}
\titlelabel{\thetitle.\quad}

\newtheorem*{thm}{Theorem}
\newtheorem{dfn}{Definition}

\title{\textbf{Axisymmetric, extremal horizons in the presence of a cosmological constant}}
\author{
	Eryk Buk\footnote{e-mail: ebuk@fuw.edu.pl}\ \ and\ Jerzy Lewandowski\footnote{e-mail: Jerzy.Lewandowski@fuw.edu.pl}\\
	\small The Faculty of Physics of the University of Warsaw\\
	\small 5 Pasteura Street, 02-093 Warsaw, Poland
}
\date{}

\begin{document}

\maketitle

\begin{abstract}
All axisymmetric solutions to the near-horizon geometry equation with a cosmological constant defined on a topological $2$-sphere were derived. The regularity conditions preventing cone singularity at the poles were accounted for. The one-to-one correspondence of the solutions with the extremal horizons in the Kerr-(anti-)de Sitter spacetimes was found. A solution corresponding to the triply degenerate horizon was identified and characterized. The solutions were also identified among the solutions to the Petrov type D equation. 
\end{abstract}

\section{Introduction} 
Einstein's equations imply constraints on the intrinsic and extrinsic geometry of the extremal Killing or isolated horizon \cite{ABL2002,LP2003,{LP2005}}. There are several such equations: the best-known constraints induced the metric tensor and the second fundamental form of the extremal horizon. That constraint was faced by H\'aji\v cek \cite{H1974}, it was written down explicitly by Isenberg and Moncrief in the case of a horizon that admits a toroidal section \cite{MI1983}, and finally it wasrediscovered, generalized to all the spacetime dimensions larger than two and systematically investigated in \cite{ABL2002,LP2003,LP2005,R2003,KL2013,PLJ2004,S2019,JK2013,KL2009CLASS,DRKLS2018,CST2017,MPS2020}. Becouse of its relevance for so-called near horizon geometry (NHG), the constraint was later named the NHG equation \cite{R2003}. In the current paper we focus on the vacuum four-dimensional spacetime case with a cosmological constant, that is, on the NHG equation imposed on a metric tensor and a one-form defined on a two-dimensional section of the extremal horizon. For $2$-manifolds of a genus higher than $0$, the general solution of the NHG equation with the cosmological constant is known \cite{DRKLS2018}. On a $2$-manifold diffeomorphic to $S_2$, on the other hand, all the axisymmetric solutions with a vanishing cosmological constant were derived and proven to correspond to the extremal Kerr Killing horizons \cite{LP2003}. They were also proven to be isolated in the possibly bigger space of all the solutions, also nonaxisymmetric \cite{CST2017}. Axisymmetric solutions of the NHG equation with a cosmological constant, on the other hand, are studied in \cite{KL2013,KL2009CLASS}. The gap we are filling in the current paper is the regularity conditions at the poles of $S_2$. We also compare our solutions with the intrinsic-extrinsic geometry of extremal horizons in the Kerr-(anti-)de Sitter spacetimes and show the one-to-one correspondence. Another new question we address in the paper is which solution to the NHG equations corresponds to a triple (rather than a double) root of the polynomial defining the horizons in those spacetimes. Finally, we compare our solutions with the axisymmetric solutions of the Petrov type D equation \cite{DRLP2018}. We identify the solutions of the Petrov type D equation that are solutions to the NHG equation.

The knowledge of all possible extremal Killing horizons was found useful for the black hole uniqueness theorems \cite{CLSH2012,CN2010,CENS2011,CT2007}. The one-to-one correspondence with the extremal Kerr horizons in the case of a vanishing cosmological constant $\Lambda$ was applied in the literature to fill some gaps concerning the extremal black holes. The generalization of the uniqueness property of the NHG equation in the $\Lambda\not=0$ case provided in the current paper shall be useful in a similar way as soon as the black hole uniqueness theorems of mathematical relativity get generalized to the asymptotically (anti) de Sitter spacetimes. That makes the research on the NHG equation still relevant and interesting for our understanding of black holes. 

 \section{Isolated, extremal horizon}
The focal point of our paper is the NHG equation. In general it can be defined on any $n$-dimensional manifold $\Delta$, endowed with metric tensor $q_{AB}$ of signature $(++\dots +)$ and one-form $\omega_A$. The NHG equation reads
\cite{LP2005}
\begin{equation} \label{eq:NHG_eq}
	\nabla_{(A}\omega_{B)} + \omega_{A}\omega_{B} - \frac{1}{2}R^{(q)}_{AB} + \frac{\Lambda}{n-2}q_{AB} = 0,
\end{equation}
where $\nabla_{A}$ is the corresponding metric and torsion-free covariant derivative, $R^{(q)}_{AB}$ is the Ricci tensor (we will mark tensors on $\Delta$ with uppercase Latin indices: $A, B, \dots$) associated with $q_{AB}$, and $\Lambda$ is a parameter. 

In order to lend this equation some physical meaning, we define the extremal isolated horizon $H$, which can be used to describe the surface of a black hole. Next we identify its section (its codimension 1 submanifold) with $\Delta$. Then the NHG equation is a constraint on geometrical data defined on $\Delta$ implied by Einstein's equations satisfied by spacetime at the extremal isolated horizon.

We give two equivalent definitions of extremal (also called \emph{degenerate} in mathematical literature), isolated horizon. Let $(M,g)$ be $(n+2)$-dimensional spacetime, that is pseudo-Riemannian manifold, with metric tensor $g$ of signature $(-+\dots +)$. The first definition is expressed in terms of the geometry of the ambient spacetime:

\begin{dfn}
	Codimension 1 hypersurface $H \subset M$ is said to be an extremal isolated horizon if it is null, and there is a vector field $N$ defined in $M$ in a neighbourhood of $H$, such that the spacetime metric $g$ and spacetime covariant derivative $\nabla$ satisfy the following conditions at $H$: 
	\begin{enumerate}[i)]
		\item
			$N$ does not vanish at any point of $H$.
		\item
			$N$ is orthogonal to $H$.
		\item
			$\left.N_{\mu}N^{\mu}\right|_{H} = 0$.
		\item
			 $\left.\mathcal{L}_{N}g\right|_{H} = 0$.
		\item
			 $\left.\left[\mathcal{L}_{N}, \nabla^{(g)}\right]\right|_{H} = 0$.
		\item
			$\nabla^{(g)}_{N}N |_{H} = 0$.
	\end{enumerate}
\end{dfn}
\bigskip

The second, equivalent definition uses intrinsic structures induced on $H$, only (we will mark tensors on $H$ with lowercase Latin indices: $a, b, \dots$):
\begin{dfn} Codimension 1 hypersurface $H \subset M$ is said to be an extremal isolated horizon if it is null, and there is a vector field $\ell$ defined on and tangent to $H$, such that the induced, degenerate metric tensor $q$ and a covariant derivative $D$, induced on $H$ by the reduction of the spacetime $\nabla$ (the reduction is well defined due to properties of $q$ assumed below) satisfy the following conditions:
	\begin{enumerate}[i)]
		\item
			$\ell$ does not vanish at any point of $H$.
		\item
			 $\ell^{a}\ell^{b}q_{ab} = 0$. 
		\item
			$\mathcal{L}_{\ell}q = 0$.
		\item
			$\left[\mathcal{L}_{\ell}, D\right] = 0$.
		\item 
			$D_\ell \ell =0.$
	\end{enumerate}
\end{dfn}
Comparing the two definitions, it is clear that $ N|_H = \ell$, $q$ is the pullback of $g$ to $H$, and owing to the condition $(iv)$ of Definition 2 the spacetime covariant derivative $\nabla$ preserves the bundle tangent to $H$ hence it reduces and induces a covariant derivative $D$. 

Given an extremal isolated horizon $(H,\ell,q,D)$, we define a rotation one-form potential $\omega$,
\begin{equation}
	D_a\ell^b =: \omega_a\ell^b.
\end{equation}

Finally, consider a codimension $2$ surface $\Delta\subset M$ that is a spacelike section of $H$, transversal to the vector field $\ell$. 
Denote by $q_{AB}$ and $\omega_A$ the data induced on $\Delta$ by $q$ and $\omega$. Now, if the spacetime metric tensor $g$ satisfies Einstein's equations (we will mark spacetime tensors with lowercase Greek indices: $\alpha, \beta, \dots$),
\begin{equation}
	R^{(g)}_{\mu\nu} - \frac{1}{2}R^{(g)}g_{\mu\nu} + \Lambda g_{\mu\nu} = 0,
\end{equation}
at the surface $H$, then Equation (\ref{eq:NHG_eq}) is defined on $\Delta$.

In the current paper we consider a two-dimensional section $\Delta$ of a three-dimensional extremal isolated horizon in four-dimensional, vacuum spacetime with cosmological constant; hence the NHG equation takes the following form:
\begin{equation} \label{eq:NHG_eq_4}
	\nabla_{(A}\omega_{B)} + \omega_{A}\omega_{B} - \frac{1}{2}R^{(q)}_{AB}+\frac{\Lambda}{2}q_{AB} = 0
	.
\end{equation}

By integrating (\ref{eq:NHG_eq_4}) over $\Delta$, and applying the Gauss-Bonnet theorem, one can derive the following equation \cite{DRKLS2018}:
\begin{equation}
	\Lambda = \frac{\int \omega^{2}\eta}{\text{Area}(\Delta)} +\frac{\int_{\Delta} K \eta}{\text{Area}(\Delta)} = 
	\frac{\int \omega^{2}\eta}{\text{Area}(\Delta)} +\frac{4 \pi}{\text{Area}(\Delta)}\Big( 1 - \text{Genus}\left(\Delta\right) \Big)
	\leq
	\frac{4 \pi}{\text{Area}(\Delta)}\Big( 1 - \text{Genus}\left(\Delta\right) \Big)
	,
\end{equation}
where $\eta$ is the area two-form of $\Delta$ and $K$ its Gaussian curvature. It is also known that all the solutions defined on a compact $2$-manifold $\Delta$ of
\begin{equation}
	\text{Genus}\left(\Delta\right) \not=0
\end{equation}
are such that \cite{DRKLS2018}
\begin{equation} \label{eq:topological_const_1}
	\omega=0,
	\quad\text{and}\quad
	K=\Lambda 
	.
\end{equation}
On the other hand,
\begin{equation} \label{eq:topological_const_2}
	\Lambda \leq \frac{4 \pi}{\text{Area}(\Delta)}
	\quad\text{for}\quad
	\Delta \cong S_{2}
	.
\end{equation}

\section{Adapted coordinates and integrating constraints}

We can use complex, null basis
\begin{equation}
	m^A m_A = \bar{m}^A\bar{m}_A = 0,
	\qquad
	m^A\bar{m}_A = 1
	,
\end{equation}
and write metric $q$ in the form
\begin{equation}
	q_{AB} = m_{A}\bar{m}_{B} + \bar{m}_{A}m_{B}.
\end{equation}
Now the covariant derivative of these basis vectors can be expressed as
\begin{equation}
	m^B \nabla_A \bar{m}_B =
	-\left(\alpha - \bar{\beta}\right)m_A
	+\left(\bar{\alpha} - \beta\right)\bar{m}_A
	= -\bar{m}^B \nabla_A m_B
	,
\end{equation}
where $\alpha$ and $\beta$ are complex functions. The rotation one-form is given by
\begin{equation}
	\omega_A = 	
	\underbrace{\left(\alpha + \bar{\beta}\right)}_{\equiv\pi}m_A
	+\underbrace{\left(\bar{\alpha} + \beta\right)}_{\equiv\bar{\pi}}\bar{m}_A
	.
\end{equation}
It will be convenient to define Gaussian curvature $K$ of our horizon, proportional to its Ricci scalar $R^{(q)}$:
\begin{equation}
	K = \frac{1}{2}R^{(q)} = 
		\delta (\alpha - \bar{\beta}) 
		+\bar{\delta}(\bar{\alpha} - \beta) 
		-2(\alpha - \bar{\beta})(\bar{\alpha} - \beta)
		,
\end{equation}
where
\begin{equation}
	\delta = m^A\partial_A,
	\qquad
	\bar{\delta} = \bar{m}^A\partial_A
	.
\end{equation}
The Ricci tensor is now given by a well-known relationship
\begin{equation}
	R^{(q)}_{AB} = \frac{1}{2}R^{(q)}q_{AB} = Kq_{AB}
	.
\end{equation}
We can use the Hodge decomposition of $\omega$ in complex functions $U$ and $B$ in the following way \cite{LP2003}:
\begin{equation} \label{eq:omega_decomposition}
	\omega = \star dU + d\log B
	.
\end{equation}
Function $U$ is defined up to an additive constant, and $B$ up to multiplicative constant. Through this decomposition, function $\pi$ is given by
\begin{equation}
	\pi = -i\bar{\delta} U + \bar{\delta}\log B
	,
\end{equation}

Using decomposition (\ref{eq:omega_decomposition}) constraints (\ref{eq:NHG_eq_4}) can be rewritten as
\begin{align}
	i\left( \bar{\delta}\delta - \delta\bar{\delta} \right)U
	+\left( \bar{\delta}\delta + \delta\bar{\delta} \right)\ln B
	-\left( \alpha - \bar{\beta}\right) \left( i\delta U + \delta\ln B \right)
	-\left( \bar{\alpha} - \beta \right) \left( -i\bar{\delta} U + \bar{\delta}\ln B \right)&\nonumber\\
	-(\delta U)^2 
	-(\bar{\delta} U)^2 
	+(\delta\ln B)^2
	+(\bar{\delta}\ln B)^2
	+2i\left( \delta U \delta\ln B - \bar{\delta}U \bar{\delta}\ln B \right)
	-K
	+\Lambda &= 0 \label{eq:constraint1and2}
	\\
	i\bar{\delta}^2U
	-\bar{\delta}^2 \ln B
	+i\bar{\delta}U\left( \alpha - \bar{\beta} \right)
	-\bar{\delta}\ln B\left( \alpha - \bar{\beta} \right)
	+(\bar{\delta}U)^2
	-(\bar{\delta} \ln B)^2
	+2i \bar{\delta}U \bar{\delta}\ln B &= 0 \label{eq:constraint3}
	,
\end{align}

Henceforth, we will be considering the NHG equation on $\Delta$ diffeomorphic to $S_2$ and for axisymmetric $q$ and $\omega$. Therefore, we introduce on $\Delta$ spherical coordinates $(\theta,\varphi)$ adapted to the axial symmetry, such that it is generated by the vector field $\partial_{\varphi}$.
The general form of axisymmetric metric $q$ on $\Delta$ is
\begin{equation}
	q_{AB}dx^{A}dx^{B} = \Sigma^2(\theta)\left( d\theta^2 + \sin^2\theta d\varphi^2 \right) .
\end{equation}

We introduce coordinate $x$ and parameter $R$ (not to be confused with Ricci scalar $R^{(g)}$ or $R^{(q)}$)
\begin{equation}
	dx = \frac{\Sigma^2(\theta)\sin\theta}{R^2}d\theta
	,
\end{equation}
where $R^2$ is defined in the following way:
\begin{equation}
	\text{Area}(\Delta)=\int_{\Delta}\Sigma^2(\theta)\sin\theta d\theta\wedge d\varphi = 2\pi R^2 (x_2- x_1)
	.
\end{equation}
Coordinate $x$ is defined up to an additive constant. We can set, say $x_1$, to an arbitrary value, and then we set $x_2$ in such a way that the area of $\Delta$ is equal to $4\pi R^2$. We will fix $x_{1}=-1$ and $x_{2}=1$ from now on.
Now the metric takes the form
\begin{equation}
	q_{AB}dx^{A}dx^{B} = R^2\left( \frac{1}{P^2(x)}dx^2 + P^2(x)d\varphi^2 \right),
	\quad
	P^2(x) = \frac{\Sigma^2(\theta) \sin^2\theta}{R^2}
\end{equation}
and the null tangent and cotangent frame, respectively, are defined by 
\begin{equation}
	m^{A}\partial_{A} = \frac{1}{\sqrt{2}R}\left( P\partial_x + i\frac{1}{P}\partial_\varphi \right),
	\qquad
	\bar{m}_{A}dx^{A} = \frac{R}{\sqrt{2}}\left( \frac{1}{P}dx - iPd\varphi \right)
	.
\end{equation}
In this basis Gaussian curvature $K$ is given by
\begin{equation} \label{eq:K}
	K = -\frac{1}{2}\frac{1}{R^2}\partial_{x}^{2}P^2
	,
\end{equation}
and one can easily calculate that
\begin{equation} \label{eq:K_1_R2}
	\frac{\int_{\Delta} K\eta}{\text{Area}(\Delta)} = \frac{1}{R^2}
	.
\end{equation}

It follows from the definition of the function $P$ that
\begin{equation} \label{eq:boundary_cond}
	P(x=\pm1) = 0.
\end{equation}	
Moreover, to avoid a conic singularity that is to ensure that the length of a circle of radius $\delta x$ about each pole is
$2\pi \delta x + o(x)$, 
\begin{equation}
\partial_{x}P^2(\pm 1) = \mp2.
\end{equation}
Notice that this is just the continuity condition on the metric tensor $q$.

Because of the axial symmetry, functions $B$ and $U$ can only depend on $x$. Real and imaginary parts of constraint (\ref{eq:constraint3}) can be written, respectively, as
\begin{align} 
	\partial_{x}^{2}B - \big(\partial_{x}U\big)^2 B &= 0 \label{eq:constraint1_re}
	\\
	\partial_{x}\big(B^2\partial_{x}U\big) &= 0 \label{eq:constraint2_re}
	.
\end{align}
while constraint (\ref{eq:constraint1and2}) takes the form
\begin{equation}
	2\frac{P P_{,x}}{R^2} \partial_{x}\log B
	+\frac{P^2}{R^2}(\partial_{x} \log B)^{2}
	+\frac{P^2}{R^2} \partial^{2}_{x}\log B
	-\frac{P^2}{R^2}(\partial_{x} U)^{2}
	+\frac{1}{2}\frac{1}{R^2}\partial^{2}_{x}P^2
	+\Lambda
	=0
\end{equation}	
Equation (\ref{eq:constraint2_re}) can be integrated to obtain
\begin{equation} \label{eq:first_int}
	U_{,x}B^2 = \tilde{\Omega}.
\end{equation}
In general $\tilde{\Omega}$ can take any value, so we will discuss the $\tilde{\Omega}\neq0$ and $\tilde{\Omega}=0$ cases separately.

\subsection{Case $\tilde{\Omega}\neq 0$}
By inserting (\ref{eq:first_int}) into (\ref{eq:constraint1_re})--(\ref{eq:constraint2_re}) and integrating, we obtain
\begin{equation}
\begin{split}
	B^{2} = B_{0}^{2}\left[ \Omega^{2} + (x - x_{0})^2 \right]\\
	U = \arctan \left( \frac{x - x_{0}}{\Omega} \right) + U_0
\end{split}
,
\end{equation}
where
\begin{equation}
	B_0,\ U_0,\ x_0 =\text{const.}\qquad\Omega = \frac{\tilde{\Omega}}{B_0^2}
	.
\end{equation}
Constraint (\ref{eq:constraint3}) can be written as
\begin{equation} \label{eq:diff_eq_for_P2}
	\partial_{x}^{2}P^2 + 
	\frac{2(x-x_{0})}{(x-x_{0})^2 + \Omega^{2}}\partial_{x}P^{2} + 
	\frac{4\Omega^{2}}{\left[ (x-x_{0})^2 + \Omega^{2} \right]^2}P^{2} = -2\Lambda R^{2} \equiv b
	.
\end{equation}
The solution to (\ref{eq:diff_eq_for_P2}) is 
\begin{equation}
\begin{aligned}
	P^{2} = \frac{1}{\Omega^2 + (x-x_{0})^{2}}\Bigg[ 
		c_{1}\left(\Omega^2 - (x-x_{0})^{2}\right) + 
		2c_{2}\Omega(x-x_{0}) \\
		+\frac{1}{2}b(x-x_{0})^2\left( \Omega^2 + \frac{1}{3}(x-x_{0})^{2} \right)
	\Bigg]
	.
\end{aligned}
\end{equation}
Applying boundary conditions (\ref{eq:boundary_cond}) is laborious, but it turns out, that both $x_0$ and $c_2$ have to vanish. This leads to
\begin{equation} \label{eq:2possibility}
	c_1 = -\frac{1}{2}b\frac{\Omega^2 + \frac{1}{3}}{\Omega^2 - 1},
	\qquad
	\Lambda R^2 = \frac{\Omega^2 - 1}{\Omega^2 - \frac{1}{3}}
	\Longleftrightarrow
	\Omega^2 = \frac{1 - \frac{1}{3}\Lambda R^2}{1 - \Lambda R^2}
	,
\end{equation}
and so we must have
\begin{equation} \label{eq:P2}
	P^2 = \left(x^2-1\right) \frac{
		\Lambda R^2 \left(
			\Lambda R^2
			-x^2 (\Lambda R^2-1)
			-5
		\right)
		+6
	}{
		\Lambda R^2+3x^2 (\Lambda R^2-1)-3
	}
	.
\end{equation}
Positivity of both this metric and $\Omega^2$ forces the following restriction:
\begin{equation}
	\Lambda R^2 \in ]-\infty,1[ \cup \{3\}
	.
\end{equation}
Now, at
\begin{equation}
	\Lambda R^2 = 3
	,
\end{equation}
 we have
\begin{equation}
	\Omega = 0 	= \tilde \Omega
\end{equation}
the case excluded in this section. Therefore, we are left with 
\begin{equation}
	\Lambda R^2 \in ]-\infty,1[ 
	,
\end{equation}
which is compatible with (\ref{eq:topological_const_2}) and (\ref{eq:K_1_R2}).
The rotation one-form is equal to
\begin{equation}
\begin{aligned}
	\omega =& \frac{x}{\Omega^2 + x^2}dx - \frac{P^2\Omega}{\Omega^2 + x^2} d\varphi =\\
	=& \frac{x(1 - \Lambda R^2)}{x^2(1 - \Lambda R^2) + (1 - \frac{1}{3}\Lambda R^2)}dx 
		-P^2\frac{\sqrt{(1 - \Lambda R^2)(1 - \frac{1}{3}\Lambda R^2)}}{x^2(1 - \Lambda R^2) + (1 - \frac{1}{3}\Lambda R^2)} d\varphi
	,
\end{aligned}
\end{equation}
where we have taken the positive root of $\Omega^2$ from (\ref{eq:2possibility}).

\subsection{Case $\tilde{\Omega}=0$}
We have the logarithm of $B$ in Equation (\ref{eq:omega_decomposition}), so we must assume 
\begin{equation}
	B>0.
\end{equation} 
By (\ref{eq:first_int}) it has to be that $U_{,x}=0$; therefore (we hope that repetitions in notation will not lead to misunderstandings),
\begin{equation}
	U = U_0,
	\qquad
	B = B_1x + B_0
	,
\end{equation}
where $U_0$, $B_{0}$, and $B_{1}$ are constants. As it will soon be apparent, we have to separate our investigation into two subcases, namely $B_1\neq 0$ and $B_1=0$.
\subsubsection{Subcase $B_1\neq 0$}
In this case constraint (\ref{eq:constraint3}) is reduced to
\begin{equation}
	 \frac{2 B_{1}}{B_1 x + B_0} \partial_{x} P^2 + \partial_{x}^{2} P^2 = b
	,
\end{equation}
which can be integrated to yield
\begin{equation}
	P^2 = 
		\frac{1}{3} b \frac{B_0}{B_1} x
		+\frac{1}{6} b x^2
		-\frac{c_1}{B_1 (B_1 x+B_0)}
		+c_2
	.
\end{equation}
As we can see, this solution is ill-defined for $B_1=0$.
After some manipulations the boundary conditions (\ref{eq:boundary_cond}) give us
\begin{equation}
	b = -6 \Leftrightarrow \Lambda R^2 = 3, \qquad
	B_0 = 0, \qquad
	c_1 = 0, \qquad
	c_2 = 1
	,
\end{equation}
which leads to following function $B$:
\begin{equation}
B=B_1 x
	,
\end{equation}
which violates the definition (\ref{eq:omega_decomposition}) of the function $B$. Hence, we exclude that case. 

\subsubsection{Subcase $B_1 = 0$}
On the other hand, if we take $B_1=0$, that means that now both the functions $U$ and $B$ are constant, and hence the constraint (\ref{eq:constraint1and2}) takes the form
\begin{equation} 
	K=\Lambda
\end{equation}
because $\omega$ vanishes. That implies
\begin{equation}
	\partial_{x}^{2} P^2 = b
	,
\end{equation}
which gives us 
\begin{equation}
	P^2 = 
		\frac{1}{2} b x^2
		+c_1 x
		+c_2
	.
\end{equation}
By applying the boundary conditions (\ref{eq:boundary_cond}) we get
\begin{equation}
	b = -2 \Leftrightarrow \Lambda R^2 = 1, \qquad
	c_1 = 1, \qquad
	c_2 = 0
	,
\end{equation}
which leads to
\begin{equation} \label{eq:lr2_1}
	P^2 = 1 - x^2
	\quad\text{and}\quad
	\omega = 0
	.
\end{equation}
These results are compatible with (\ref{eq:topological_const_1}) and (\ref{eq:K_1_R2}).

\section{Comparison with earlier results}
\subsection{Review article by H. K. Kunduri and J. Lucietti}
In a review article \cite{KL2013} the authors calculated the solution to the same problem:
\begin{equation} \label{eq:KLresult}
	\frac{P^2}{R^2} = 
		\frac{4\beta}{
			4k^2
			+\beta^2 x^2
		}\left[
			-\frac{\beta\Lambda x^4}{12}
			+\left( 
				A_0 
				-\frac{2\Lambda k^2}{\beta}
			\right) x^2
			-\frac{4 k^2}{\beta^2}\left( 
				A_0 
				-\frac{\Lambda k^2}{\beta}
			\right)
		\right]
\end{equation}
where $\beta$, $k$, and $A_0$ are real constants. However, they did not consider the problem of conical singularity, thus leaving an independent parameter. By applying boundary conditions (\ref{eq:boundary_cond}) we get relationships between constants in (\ref{eq:KLresult}) and our parameters:
\begin{equation} 
	\frac{4k^2}{\beta^2} = \frac{1 - \frac{1}{3}\Lambda R^2}{1 - \Lambda R^2}
	,
	\qquad
	\frac{2 A_0}{\beta} = \Lambda \left( \frac{1 - \frac{2}{3}\Lambda R^2}{1 - \Lambda R^2} - \frac{1}{\Lambda R^2} \right)
	.
\end{equation}
Thus the following equation constraints parameters in (\ref{eq:KLresult})
\begin{equation} \label{eq:KL_constr}
	A_0 = \Lambda \beta\frac{
		 \left(1 - 3\frac{16 k^4}{\beta^4} +6 \frac{4 k^2}{\beta^2}\right)
	}{
		12 \left(1 - \frac{4 k^2}{\beta^2}\right)
	}
	.
\end{equation}
Furthermore, as the authors suggest in \cite{KL2009ADS}, one of the constants $\beta$, $k$, or $A_0$ can be eliminated, via rescaling. This and Equation (\ref{eq:KL_constr}) reduce the number of parameters to one, just as in our case.\\

\subsection{Solution to Petrov type D equation}
	In \cite{DRLP2018} an equation is considered that is an integrability condition for our NHG equation (every metric satisfying the latter also satisfies the former; see \cite{DRKLS2018}), namely
	
		\begin{equation} \label{eq:type_D}
		\left(
			\bar{\delta} + \alpha - \bar{\beta}
		\right)
		\bar{\delta}\Psi_{2}^{-\frac{1}{3}}
		,
	\end{equation}
	where the complex valued function $\Psi_{2}$ can be expressed via functions defined on $\Delta$,
	\begin{equation}
		\Psi_2 = -\frac{1}{2}\left(K + i\mathcal{O}\right) + \frac{\Lambda}{6}
		,
	\end{equation}
	where
	\begin{equation}
		\mathcal{O} = -\Big[
			\delta\bar{\delta} 
			+\bar{\delta}\delta 
			-(\alpha - \bar{\beta})\delta 
			-(\bar{\alpha} - \beta)\bar{\delta}
		\Big]U
		.
	\end{equation}
	All the axisymmetric solutions are derived. The metric calculated in \cite{DRLP2018} has the form
	\begin{equation} \label{eq:type_D_sol}
		P^2 = (1 - x^2) - \frac{1}{1-\frac{\gamma\Lambda}{6}}\frac{(1-x^2)^2}{x^2+\eta^2}
		,
	\end{equation}
	where parameters $\gamma$ and $\eta$ are real. It was not analyzed which of those solutions correspond to solutions of the NHG equation, but it is done in this section.\\
	If we substitute 
	\begin{equation}
		\eta^2 = \frac{1 - \frac{1}{3}\Lambda R^2}{1 - \Lambda R^2},
		\qquad
		\gamma = \frac{6 R^2}{\Lambda R^2 - 3}
		,
	\end{equation}
	then both (\ref{eq:type_D_sol}) and (\ref{eq:P2}) agree.
	Solution (\ref{eq:lr2_1}), corresponding to $\Lambda R^2 = 1$, is also the same. 
	Therefore we have reproduced results from \cite{DRLP2018} for
	\begin{equation}
		\Lambda R^2 \in \left]-\infty, 1\right]
		.
	\end{equation}
	This is the way the solutions calculated in the current paper are sitting among the solutions found in \cite{DRLP2018}.

	Equation (\ref{eq:type_D}) has also applications to nonextremal isolated horizons. Recall, that the Weyl tensor of four-dimensional spacetime has principal directions, $4$-distinct and null in the generic case. At an isolated horizon, two of the principal null directions come together and are tangent to the horizon. That makes the Weyl tensor to be of the Petrov type II. If the remaining two null directions coincide as well, then the Weyl tensor is said to be of the Petrov type D. The remaining possibilities are excluded at isolated horizons except when the Weyl tensor vanishes \cite{DRLP2018a}. The assumption that the Weyl tensor is of the Petrov type D at a nonextremal isolated horizon leads to Equation (\ref{eq:type_D}) \cite{DRLP2018a} (and earlier in \cite{LP2002}). 

\section{Embedding in Kerr-(anti)de Sitter spacetime}
The Kerr-(anti) de Sitter spacetimes are the Petrov type D vacuum solutions to Einstein's equations with a cosmological constant and set a family parametrized by constants $m$, $a$, and $\Lambda$. They contain Killing horizons that are automatically our isolated horizons. For special values of the parameters, two of the generically distinct horizons coincide. Then, the resulting horizon is extremal also in the meaning of our definition of the extremal isolated horizon. Our aim now is to compare the axially symmetric solutions to the NHG equation derived in the current paper with the data defined on a section of the Kerr-(anti)de Sitter extremal horizons (Section \ref{KdS}), which will lead us to the uniqueness theorem (Section \ref{Uniqueness}). 

\subsection{Extremal Kerr-de Sitter spacetime}\label{KdS}
Kerr-de Sitter spacetime has a metric of well-known form
\begin{equation}
	g = 
		-\frac{\Delta_r}{\chi^2 \rho^2}\left(
			dt - a \sin^2\theta d\varphi
		\right)^2
		+\frac{\Delta_\theta\sin^2\theta}{\chi^2 \rho^2}\left(
			adt - (r^2 + a^2)d\varphi
		\right)^2
		+\rho^2\left(
			\frac{dr^2}{\Delta_r} + \frac{d\theta^2}{\Delta_\theta}
		\right)
\end{equation}
where:
\begin{equation}
\begin{aligned}
	\rho^2 &= r^2 + a^2 \cos^2\theta\\
	\Delta_\theta &= 1 + \frac{1}{3}\Lambda a^2 \cos^2\theta\\
	\chi &= 1 + \frac{1}{3}\Lambda a^2\\
	\Delta_r &= (r^2 + a^2)\left( 1 - \frac{1}{3}\Lambda r^2 \right) - 2Mr
\end{aligned}
\end{equation}
The vanishing of polynomial $\Delta_r$ discerns values of $r=r_0$ for which Killing vectors form a horizon. Extremal horizons correspond to its multiple roots.
By simple manipulations it can be shown, that both metric $P^2$ and quantity $R^2$ can be expressed \cite{DRLP2018} as
\begin{equation} \label{eq:P2andR2}
	P^2 = 
		(1 - x^2)
		\frac{1 + \frac{1}{3}\Lambda a^2 x^2}{1 + \frac{1}{3}\Lambda a^2}
		\frac{r_0^2 + a^2}{r_0^2 + a^2 x^2}
		,
	\qquad
	R^2 = \frac{r_0^2 + a^2}{1 + \frac{1}{3}\Lambda a^2}
	,
\end{equation}
and coordinate $x$ takes the form
\begin{equation}
	x = -\cos\theta
	.
\end{equation}
Extremal horizons can be found by equating the discriminant of $\Delta_r$ to zero. The discriminant can be expressed only in terms of $a^2$, $\Lambda$, and $R^2$ by calculating $M$ from $\Delta_r=0$ and eliminating $r_0$ using the definition of $R^2$ in (\ref{eq:P2andR2}):
\begin{equation} \label{eq:discr}
\begin{split}
	\frac{\Lambda}{{a^2 (\Lambda R^2-3)+3R^2}}
	\left[a^2 \Lambda +3\right] 
	\Big[a^2 (\Lambda R^2-3) (\Lambda R^2-2)+3 R^2 (\Lambda R^2-1)\Big]\cdot\\
	\cdot (\Lambda R^2-3)\Big[
		a^4 \Lambda ^3 R^4 (\Lambda R^2+1)
		+6 a^2 (\Lambda R^2 (\Lambda R^2-6) (\Lambda R^2+3)+54)\\
		+9 R^2 (\Lambda R^2-4) (\Lambda R^2-3)
 	\Big]
	=0
	.
\end{split}
\end{equation}
We must also account for the following conditions:
\begin{equation} \label{eq:parameter_conds}
	a^2 \geq 0,
	\quad
	R^2 > 0,
	\quad
	r_0^2 \geq 0,
	\quad
	P^2 \geq 0,
	\quad
	M\geq0
	.
\end{equation}
One can easily see that $a^2\Lambda = -3$ would make metric $g$ ill-defined, similar to $\Lambda R^2 = 3$. Setting $\Lambda = 0$ makes the discriminant vanish, but it will be contained in further results.\\
It turns out that if we apply (\ref{eq:parameter_conds}), then the only possible solution for $a^2$ is
\begin{equation} \label{eq:a2_KdS}
	a^2 = \frac{
		3 R^2 (1-\Lambda R^2)
	}{
		(3-\Lambda R^2) (2-\Lambda R^2)
	}
	,
\end{equation}
which, together with (\ref{eq:parameter_conds}), restricts our parameters in the following way:
\begin{equation}
	\Lambda R^2 \in ]-\infty,1]\cup]2,3[
	.
\end{equation}
If we take the form of $a^2$ from (\ref{eq:a2_KdS}), then mass $M$ is given by
\begin{equation} \label{eq:M_KdS}
		M = \frac{2}{3} \sqrt{\frac{R^2}{2-\Lambda R^2 }} \frac{(3-2 \Lambda R^2 )^2 }{(2-\Lambda R^2) (3-\Lambda R^2)}
		,
\end{equation}
which precludes $\Lambda R^2 > 2$ from the allowed range of parameters. Our metric must take the following form:
\begin{equation} \label{eq:P2_KdS}
	P^2 = \left(x^2-1\right) \frac{
		\Lambda R^2 \left(
			\Lambda R^2
			-x^2 (\Lambda R^2-1)
			-5
		\right)
		+6
	}{
		\Lambda R^2+3x^2 (\Lambda R^2-1)-3
	}
	,
\end{equation}
where
\begin{equation}
	\Lambda R^2 \in ]-\infty,1]
\end{equation}

Function $P^2$ in (\ref{eq:P2_KdS}) has been calculated independently, from (\ref{eq:P2}), yet takes identical form. Thus every extremal Kerr-de Sitter metric must take the form (\ref{eq:P2}), with parameters $M$ and $a^2$, given by (\ref{eq:M_KdS}) and (\ref{eq:a2_KdS}) respectively, and with $\Lambda R^2 \leq 1$.

\subsection{The uniqueness of the axisymmetric extremal isolated horizons} \label{Uniqueness}
An important conclusion of the previous subsection is the following uniqueness theorem:
\begin{thm}
	Suppose $H$ is an isolated horizon in four-dimensional spacetime, that satisfies the vacuum Einstein equations with a (possibly vanishing) cosmological constant; suppose also that $H$ satisfies each of the following conditions:
	\begin{itemize}
		\item 
			$H$ admits a two-dimensional, spacelike section $\Delta$, diffeomorphic to $S_2$,
		\item 
			the metric tensor $q_{AB}dx^{A}dx^{B}$ induced on $\Delta$ and the pullback $\omega_A dx^A$ to $\Delta$ of the rotation one-form potential are axially symmetric. 
	\end{itemize}
	Then, $(q_{AB}, \omega_{A})$ coincides with the data defined on a section of an extremal horizon in one of the Kerr-(anti-)de Sitter spacetimes.
\end{thm}

\subsection{Doubly extremal horizon}
We will now investigate when the horizon of Kerr-de Sitter is extremal, that is, when roots of $\Delta_r$ merge. In particular, we want to find out, when three roots merge and horizon is doubly extremal.
To describe the dependence of roots of $\Delta_r$ on its coefficients, we will use rules elucidated in \cite{Rees1922} and the well-known Descartes rule of signs. It will allow us to find a kind (real or imaginary, positive or negative) of a root and its multiplicity in the range of our parameters. This information is contained in Table \ref{tb:roots}.

\begin{table}[hb!]
\caption{Dependence of roots of $\Delta_r$ on parameters.}
\label{tb:roots}
\begin{center}
\begin{tabular}{c|c|c}
	Case & Parameter ranges & Number and type of roots \\ \hline
	$(i)$ & $\Lambda R^2 < 0$, $R^2\neq0$ & 1 real, positive, double; 2 imaginary \\ \hline
	$(ii)$ & $\Lambda R^2 = 0$ & 1 real, positive, double \\ \hline
	$(iii)$ & $\Lambda R^2 \in ]0,1] \backslash \left\{\frac{3 - \sqrt{3}}{2}\right\}$ & all real; 1 positive, double; 1 positive and 1 negative \\ \hline
	$(iv)$ & $\Lambda R^2 = \frac{3 - \sqrt{3}}{2}$ & both real; 1 positive, triple; 1 negative
\end{tabular}
\end{center}
\end{table}

Cases $(iii)$ and $(iv)$ are particularly interesting, because we have three roots, two of them merged, that will all become equal for $\Lambda R^2 = (3 - \sqrt{3})/2$. We will now write down their explicit forms:
\begin{equation}
\begin{aligned}
	r_1 &= \frac{-\sqrt{A} + \sqrt{B + C\sqrt{A}}}{\sqrt{2}}, \\
	r_2 &= \frac{\sqrt{A} - \sqrt{B - C\sqrt{A}}}{\sqrt{2}}, \\
	r_3 &= \frac{\sqrt{A} + \sqrt{B - C\sqrt{A}}}{\sqrt{2}}; \\
\end{aligned}
\end{equation}
where
\begin{equation}
\begin{aligned}
	A &= \frac{(2 \Lambda R^2 -3)^2}{\Lambda \left(\Lambda R^2 - 2 \right) \left(\Lambda R^2 - 3 \right)}, 
	\\
	B &= \frac{3}{ 2\Lambda }+\frac{R^{2} (\Lambda R^{2}+3)}{2(\Lambda R^{2}-3) (\Lambda R^{2}-2)},
	\\
	C &= 2 \sqrt{2} \sqrt{\frac{R^2}{2-\Lambda R^2 }}.
\end{aligned}
\end{equation}
As we can easily calculate, for positive $\Lambda R^2$ the following holds:
\begin{equation}
	r_1 = r_2
	\quad\text{for}\quad
	\Lambda R^2 > 0
\end{equation}
and
\begin{equation}
	r_1 = r_2 = r_3 = R\sqrt{ \sqrt{3} - 1 }
	\quad\text{for}\quad
	\Lambda R^2 = \frac{3 - \sqrt{3}}{2}
	.
\end{equation}

The first two positive roots are always merged, and the equality of all three of them reduces to case $(iv)$. In case $(ii)$ we get
\begin{equation}
	a^2 = M^2 = r^2 = \frac{R^2}{2}
	,
\end{equation}
which agrees with the results for the Kerr metric.

\section{Summary}
We have studied axially symmetric solutions $(q,\omega)$ to the near horizon geometry equation with cosmological constant (\ref{eq:NHG_eq_4}) on two-dimensional manifold $\Delta$ diffeomorphic to sphere. We have shown that every solution is determined by the values of cosmological constant $\Lambda$, and total area $4\pi R^2$, bounded by the following condition:
\begin{equation}
	-\infty < \Lambda R^2 \le 1
	.
\end{equation}
The metric $q$ is of the general form
\begin{equation}
	q_{AB}dx^{A}dx^{B} = R^2\left( \frac{1}{P^2(x)}dx^2 + P^2(x)d\varphi^2 \right),
\end{equation}
where the function $[-1,1]\ni x\mapsto P(x)$ is defined as follows:
\begin{equation}
	P^2(x) = \left(x^2-1\right) \frac{
		\Lambda R^2 \left(
			\Lambda R^2
			-x^2 (\Lambda R^2-1)
			-5
		\right)
		+6
	}{
		\Lambda R^2+3x^2 (\Lambda R^2-1)-3
	}
	,
\end{equation}
	 while the rotation one-form takes the form
	\begin{equation}
	\begin{aligned}
		&\omega = \frac{x(1 - \Lambda R^2)}{x^2(1 - \Lambda R^2) + (1 - \frac{1}{3}\Lambda R^2)}dx 
		\pm P^2\frac{\sqrt{(1 - \Lambda R^2)(1 - \frac{1}{3}\Lambda R^2)}}{x^2(1 - \Lambda R^2) + (1 - \frac{1}{3}\Lambda R^2)} d\varphi
		\quad\text{for}\quad
		\Lambda R^2 < 1,\\
		&\omega = 0 \quad\text{for}\quad \Lambda R^2 = 1. 
	\end{aligned}
	\end{equation}
They are defined on $[-1,1]\times S_1$ endowed with coordinates $(x,\varphi)$ and the regularity conditions at the poles $x=\pm 1$ make them continuous on $S_2$.
	
We have compared our results with those contained in the review by Kunduri and Lucietti \cite{KL2013}. In particular, we have transformed our parameters to those used in the review and solved explicitly the restrictions following from our regularity assumptions: bounded $\omega$ and continuous $q$ of the signature $(++)$. 

It is known that the Petrov type D equation \cite{DRLP2018a} is an integrability condition for the NHG equation. Hence the axisymmetric solutions to the Petrov type D equation found in \cite{DRLP2018} contain solutions to the NHG equation; however they were not identified. We have filled that gap in the current paper. 
 
Furthermore, for every solution $(q,\omega)$ we have determined the corresponding extremal Kerr-(anti)de Sitter spacetime (see Theorem in Section \ref{Uniqueness}): given 
$\Delta$, $q$, and $\omega$ were embedded in the extremal horizon in that spacetime, such that $q$ coincides with the pullback of the spacetime metric tensor and $\omega$ coincides with the pullback of the spacetime rotation one-form potential. In particular, we have identified those solutions $(q,\omega)$ of the NHG equation that correspond to a triple root of the polynomial, whose roots define the horizons of the Kerr-(anti)de Sitter spacetimes.

The assumptions on the axial symmetry and $\Lambda$ vacuum made in this paper can be relaxed in future research. On the one hand, coupling with the Maxwell or even Yang-Mills fields should lead to a generalization of our result as it is the case when $\Lambda=0$ \cite{LP2003,KL2013}. The existence of nonaxially symmetric solutions defined on a topological $2$-sphere, on the other hand, is a hard problem that has been approached; however only partial results are known \cite{CST2017}. Another possibility is a generalization to higher dimensions.

\begin{center}\textbf{Acknowledgements}\end{center}
We thank Maciej Kolanowski and Pawe\l{} Sobiecki for information about the parameters of the extremal Kerr-(anti)de Sitter spacetimes. Jerzy Lewandowski was supported by the Polish National Science Centre, Grant No. 2017/27/B/ST2/02806.

\end{document}